\documentclass[aps,prr,twocolumn,showpacs,letterpaper,superscriptaddress,longbibliography]{revtex4-1}

\usepackage[colorlinks=true, citecolor=blue, linkcolor=blue, urlcolor=blue]{hyperref}
\usepackage{graphicx,dcolumn,longtable,epsfig}
\usepackage[usenames]{color}
\usepackage{amssymb}
\usepackage{amsmath}
\usepackage{bm}
\usepackage{footnote}
\usepackage{float}
\usepackage{subfigure}
\usepackage{color}

\usepackage{ulem}
\usepackage[T1]{fontenc}

\newcommand{\be}{\begin{equation}}
\newcommand{\ee}{\end{equation}}

\def\bea{\begin{eqnarray}}
\def\eea{\end{eqnarray}}

\begin{document}

\title{Superfluid transition of bond bipolarons with long-range Coulomb repulsion in two dimensions}

\author{Chao Zhang}
\email{chaozhang@ahnu.edu.cn}
\affiliation{Department of Physics, Anhui Normal University, Wuhu, Anhui 241000, China}

\begin{abstract}

Using numerically exact diagrammatic Monte Carlo simulations in the two-electron (single-bipolaron) sector, we explore the impact of long-range Coulomb repulsion on the dilute-limit Berezinskii--Kosterlitz--Thouless (BKT) transition temperature $T_c$ of bipolarons on a two-dimensional square lattice.
We study the bond Su--Schrieffer--Heeger model, in which bond phonons modulate the electron hopping.
In the absence of long-range repulsion, this model was shown to support small, light bipolarons with a comparatively high transition temperature \cite{PhysRevX.13.011010}.
Here we find that long-range Coulomb repulsion suppresses the optimal $T_c$ but leaves it appreciable over a broad parameter window, including the adiabatic regime $\omega/t=0.5$ at a representative Coulomb strength $V=U/10$ (with $U$ the on-site repulsion).
Our results provide controlled single-bipolaron inputs for dilute-limit $T_c$ estimates in the presence of long-range repulsion.

\end{abstract}

\maketitle

\section{Introduction}

Electron–phonon coupling is a central ingredient of condensed-matter physics and underlies a wide range of phenomena, from charge-carrier dressing and transport anomalies to the formation of polaronic and bipolaronic quasiparticles~\cite{Landau33,Frohlich50,Feynman:1955du,Schultz:1959el,Holstein59,Alexandrov:1999fy,Holstein2000, 6127-phps, PhysRevB.108.245127, PhysRevB.107.L121109}. When two electrons bind into a bipolaron via phonon-mediated attraction, a superconducting state, or more generally a superfluid state of composite bosons, may emerge. A key requirement for such a scenario is that the bipolaron remain sufficiently light and compact: a large effective mass or an extended pair size strongly suppresses the superfluid stiffness and thereby lowers the transition temperature, even when pairing itself is robust. In two dimensions, where the transition is of Berezinskii–Kosterlitz–Thouless type, and in three dimensions, where finite-temperature condensation is possible, the bipolaron mass and size are therefore central quantities in assessing the achievable transition temperature.

Two prototypical electron--phonon coupling mechanisms illustrate the challenge.
In the Holstein model, phonons couple locally to the electronic density, and extensive work has shown that both polaron and bipolaron effective masses typically grow rapidly (often exponentially) with coupling strength, which severely limits the transition temperature in the strong-coupling regime~\cite{PhysRevB.55.R8634,PhysRevLett.84.3153,PhysRevB.69.245111,PhysRevLett.105.266605}.
In contrast, in the bond Su--Schrieffer--Heeger (SSH) model the lattice degrees of freedom modulate the electron hopping amplitudes.
This SSH coupling can generate polarons whose effective mass remains comparatively light even at strong coupling, a property that has stimulated substantial interest in SSH polaron physics~\cite{PhysRevLett.121.247001,PhysRevB.109.L220502, PhysRevLett.105.266605,PhysRevB.104.035143,ckbn-jp9t,PhysRevB.104.L140307}.
These observations naturally raise the prospect that SSH-type coupling may also support light bipolarons, and therefore an enhanced superfluid transition temperature, in regimes where Holstein-based bipolarons become prohibitively heavy.
Moreover, the persistence of light quasiparticles under additional material-motivated ingredients (e.g., phonon dispersion or mixed couplings) further motivates assessing the robustness of the SSH bipolaron route~\cite{PhysRevB.108.075156, PhysRevB.109.165119, PhysRevB.111.184513, arXivchao2025, PhysRevB.111.134504}.

In the absence of long-range Coulomb repulsion, the bond SSH model was shown to form small, light bipolarons that yield a comparatively high transition temperature in two dimensions~\cite{PhysRevX.13.011010}.
In three dimensions, it was further argued that a high transition temperature can persist even when long-range Coulomb repulsion is included, reaching values comparable to or exceeding common McMillan-type adiabatic estimates for phonon-mediated superconductivity~\cite{PhysRevB.108.L220502}.
These results motivate a focused question that is especially acute in low dimensions: to what extent does long-range Coulomb repulsion compromise the high transition temperature of SSH bipolarons in two dimensions?
More specifically, it is important to determine (i) how the maximal attainable transition temperature is suppressed by Coulomb repulsion, (ii) whether the optimal coupling window shifts, and (iii) what adiabatic parameter regimes (small phonon frequency) can still support a sizable transition temperature.
Answering these questions is essential for establishing the SSH bipolaron mechanism as a viable low-dimensional route to bipolaron superconductivity, where long-range repulsion is unavoidable and can strongly reshape both pairing and phase coherence.

In this paper, we address these questions by studying the bond SSH model on the two-dimensional square lattice in the presence of long-range Coulomb repulsion. We employ numerically exact diagrammatic Monte Carlo simulations in the two-electron sector, combining a path-integral representation for the electrons with real-space diagrammatic sampling of phonons~\cite{PhysRevB.105.L020501}. From the single-bipolaron dispersion and real-space correlations, we extract the effective mass and spatial extent of the pair, and use these controlled single-bipolaron inputs to construct an optimized dilute-limit estimate of the BKT transition temperature $T_c$, rather than the temperature from an explicit finite-density calculation. We find that long-range repulsion generally suppresses the optimal transition temperature, but leaves it sizable over a broad parameter window. In particular, sizable estimated values persist in the adiabatic regime $\omega/t=0.5$ at $V=U/10$, reflecting the survival of relatively light and compact bipolarons in this regime.

The rest of this paper is organized as follows.
In Sec.~\ref{sec:sec2}, we present the Hamiltonian of the bond SSH model with long-range Coulomb repulsion.
In Sec.~\ref{sec:sec3}, we describe the simulation method and the single-bipolaron observables.
In Sec.~\ref{sec:sec4}, we outline the construction of the dilute-limit BKT transition-temperature estimate.
In Sec.~\ref{sec:sec5}, we present and discuss our results, and Sec.~\ref{sec:sec6} concludes the paper.

\section{Hamiltonian}
\label{sec:sec2}

We study electrons on a two-dimensional square lattice coupled to bond phonons of Su--Schrieffer--Heeger (SSH) type. 
In this model, an independent Einstein oscillator is associated with each nearest-neighbor bond $\langle ij\rangle$, and its displacement modulates the kinetic energy of the electrons.

The Hamiltonian reads~\cite{PhysRevLett.42.1698,PhysRevLett.25.919,PhysRevB.5.932,PhysRevB.5.941}
\begin{equation}
\begin{aligned}
H &= H_{e}+H_{\rm ph}+H_{\rm int},\\
H_{e} = & -t\sum_{\langle ij\rangle,\sigma}\left(c^{\dagger}_{j\sigma}c_{i\sigma}+{\rm H.c.}\right)
+U\sum_{i}n_{i\uparrow}n_{i\downarrow} \\
&+\frac12\sum_{i\neq j}V_{ij}\,n_i n_j,\\
H_{\rm ph} &= \omega\sum_{\langle ij\rangle}\left(b^{\dagger}_{\langle ij\rangle}b_{\langle ij\rangle}+\frac12\right),\\
H_{\rm int} &= g\sum_{\langle ij\rangle,\sigma}\left(c^{\dagger}_{j\sigma}c_{i\sigma}+{\rm H.c.}\right)X_{\langle ij\rangle},\\
X_{\langle ij\rangle} &= b_{\langle ij\rangle}+b^{\dagger}_{\langle ij\rangle}.
\end{aligned}
\label{Eq1}
\end{equation}
Here $c^{\dagger}_{i\sigma}$ ($c_{i\sigma}$) creates (annihilates) an electron with spin $\sigma\in\{\uparrow,\downarrow\}$ on site $i$, and
$n_{i\sigma}=c^{\dagger}_{i\sigma}c_{i\sigma}$ with $n_i=n_{i\uparrow}+n_{i\downarrow}$.
The operator $b^{\dagger}_{\langle ij\rangle}$ ($b_{\langle ij\rangle}$) creates (annihilates) a phonon on the bond $\langle ij\rangle$; correspondingly, $X_{\langle ij\rangle}$ is the dimensionless bond displacement.
The hopping amplitude is $t$, and $U$ is the on-site Hubbard repulsion (unless stated otherwise, we fix $U/t=8.0$).
The long-range Coulomb repulsion is parameterized as
\begin{equation}
V_{ij}= V\,\frac{a}{|{\bf r}_i-{\bf r}_j|}\qquad (i\neq j),
\end{equation}
where $V$ sets the interaction strength, ${\bf r}_i$ is the position of site $i$, and $a$ is the lattice constant (set to $a=1$ below).
$\omega$ is the phonon frequency and $g$ is the electron-phonon coupling strength.

The physics is conveniently organized by (i) a dimensionless SSH coupling $\lambda$ and (ii) the adiabaticity ratio $\omega/t$.
Following Ref.~\cite{PhysRevX.13.011010}, we define
\begin{equation}
\lambda \equiv \frac{g^2}{D\,t\,\omega}\qquad (D=2),
\label{lambda}
\end{equation}
so that $\lambda$ is dimensionless and can be compared across parameters.
We focus on the adiabatic regimes $\omega/t\le 1$, where the phonon degree of freedom is considered comparable or slow with respect to the electronic motion, and present results for $\omega/t=1.0$ and $0.5$ (the bare bandwidth in 2D is $W=8t$).

For smaller $\omega/t$ and/or stronger Coulomb repulsion, the bipolaron dispersion becomes increasingly flat and the effective mass grows rapidly, which makes an accurate determination of $m_{\rm BP}^*$ numerically demanding.
In this regime the relevant energy differences at the smallest momenta are tiny, so longer imaginary-time projection and substantially improved statistics are required to resolve the curvature of $E_{\rm BP}(\mathbf{k})$ reliably.
Accordingly, we focus on $\omega/t=1.0$ and $0.5$ and on Coulomb strengths up to $V=U/4$.

Finally, unless stated otherwise we set $U/t=8.0$.
This choice is motivated by Ref.~\cite{PhysRevX.13.011010}, which found a dome-like dependence of the optimized transition temperature on $U/t$ in two dimensions, with the maximum occurring near $U/t\simeq 8.0$.
\par

\begin{figure}[t]
\includegraphics[width=0.48\textwidth]{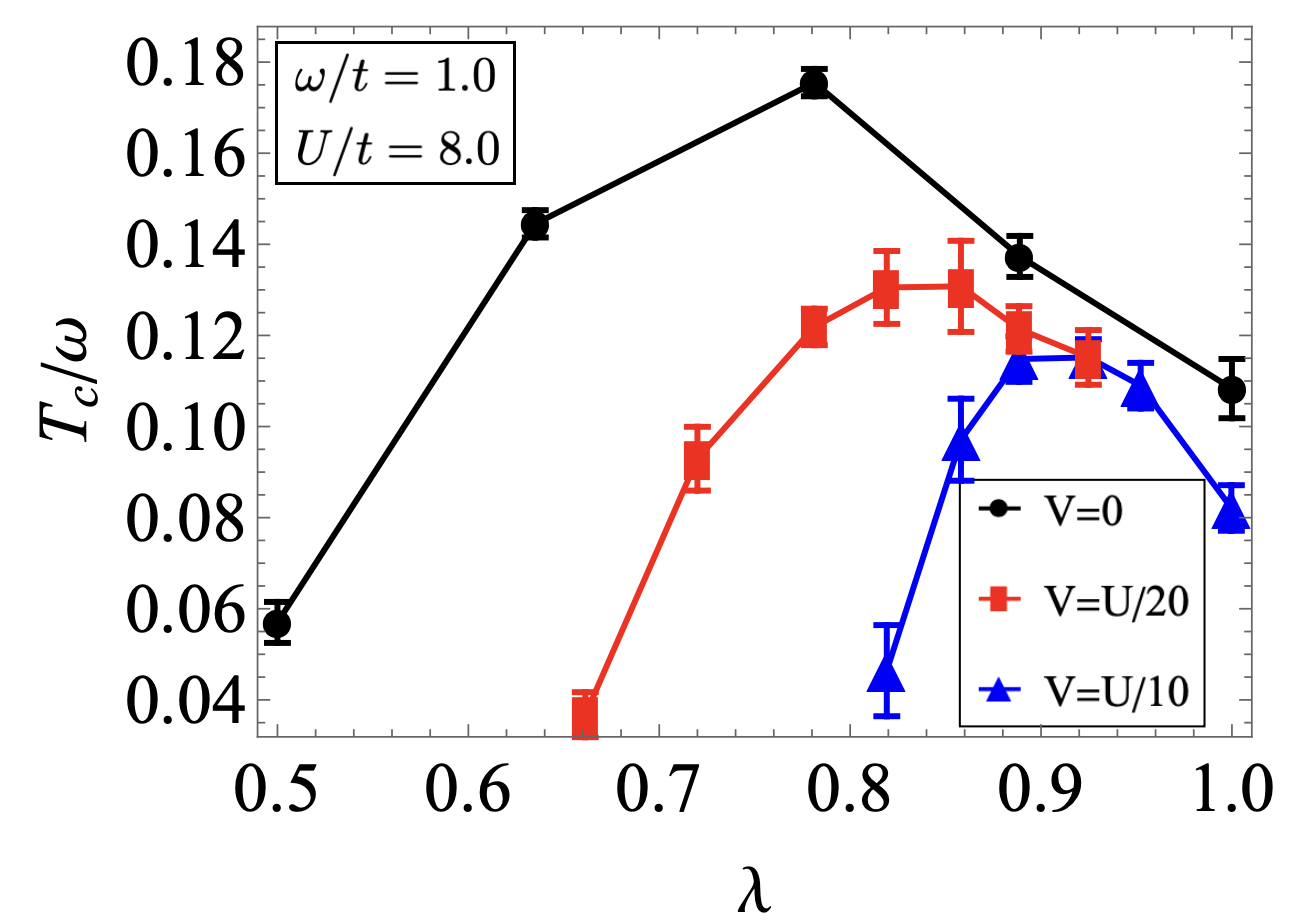}
\caption{Dilute-limit BKT estimates of $T_c/\omega$ for bond (SSH) bipolarons as a function of the dimensionless coupling $\lambda$ [defined in Eq.~\ref{lambda}], for Coulomb strengths $V=0$ (black dots), $V=U/20$ (red squares), and $V=U/10$ (blue triangles), at fixed $\omega/t=1.0$ and $U/t=8.0$.
The estimates are obtained from Eq.~\ref{Eq2} using diagrammatic Monte Carlo results for the single-bipolaron effective mass $m^{*}_{\mathrm{BP}}$ and mean-square radius $R^2_{\mathrm{BP}}$ in the two-electron sector.
Error bars denote one standard deviation (statistical).
}
\label{FIG1}
\end{figure}

\section{Method}
\label{sec:sec3}

We employ a numerically exact diagrammatic Monte Carlo approach based on a path-integral representation for the electronic sector combined with real-space diagrammatic sampling of the phonon degrees of freedom~\cite{PhysRevB.105.L020501}.
All results reported in this work are obtained in the two-electron (single-bipolaron) sector of Hamiltonian~\eqref{Eq1}, yielding unbiased estimates within statistical uncertainties.

We simulate an $L\times L$ square lattice with open boundary conditions (OBC) and use $L=140$ as our default size.
Unless stated otherwise, all figures correspond to $L=140$; finite-size effects are tested and found to be smaller than the statistical uncertainties in the parameter window shown.

The bipolaron dispersion $E_{\mathrm{BP}}(\mathbf{k})$ and internal structure are extracted from the pair Green's function
\begin{equation}
G_{\mathrm{BP}}(\mathbf{R},\mathbf{r},\tau)=
\left\langle
c_{\mathbf{r}_1,\uparrow}(\tau)\,c_{\mathbf{r}_2,\downarrow}(\tau)\,
c^{\dagger}_{0,\downarrow}(0)\,c^{\dagger}_{0,\uparrow}(0)
\right\rangle ,
\label{G2}
\end{equation}
where $\mathbf{R}=(\mathbf{r}_1+\mathbf{r}_2)/2$ and $\mathbf{r}=\mathbf{r}_1-\mathbf{r}_2$ are the center-of-mass and relative coordinates, respectively. 

In the asymptotic $\tau\to\infty$ limit, the correlator projects onto the lowest-energy state in the corresponding momentum sector; for a stable (non-decaying) bipolaron,
\begin{equation}
G_{\mathrm{BP}}(\mathbf{k},\mathbf{p}=0,\tau)\propto
e^{-\left[E_{\mathrm{BP}}(\mathbf{k})-\mu\right]\tau},
\qquad (\tau\to\infty),
\label{Green}
\end{equation}
so that $E_{\mathrm{BP}}(\mathbf{k})$ is obtained from the exponential decay rate. The
chemical potential $\mu$ here is added to shift energies and
control the rate of the exponential decays. It is used for
computational convenience and the extracted dispersion is independent of the specific choice of $\mu$.

In this work we measure the bipolaron binding energy, effective mass, and size.

\textit{Binding energy:}
\begin{equation}
\Delta_{\mathrm{BP}}(\mathbf{k}=0)=2E_{\mathrm{p}}(\mathbf{k}=0)-E_{\mathrm{BP}}(\mathbf{k}=0),
\end{equation}
where $E_{\mathrm{p}}(\mathbf{k}=0)$ is the single-polaron ground state energy.

\textit{Effective mass:}
we extract the bipolaron effective mass from the small-$\mathbf{k}$ curvature of the dispersion,
\begin{equation}
m_{\mathrm{BP}}^{*}/m_0=2t/ \frac{\partial^2 E_{\mathrm{BP}}(\mathbf{k})}{\partial \mathbf{k}^2},
\end{equation}
implemented in practice by fitting $E_{\mathrm{BP}}(\mathbf{k})$ at the smallest accessible momenta. Here, the mass of two free electrons
$m_0 = 2m_e = 1/t$.

\textit{Mean-square radius:}
to characterize the bound-state structure we collect statistics for the relative separation $\mathbf{r}$ of the two electrons in the middle of the large-$\tau$ trajectory.
Let $P(\mathbf{r})$ be the corresponding ground-state probability distribution. We define
\begin{equation}
R^2_{\mathrm{BP}}=\left\langle\left(\frac{r}{2}\right)^2\right\rangle
=\sum_{\mathbf{r}}\left(\frac{r}{2}\right)^2 P(\mathbf{r}),
\end{equation}
which corresponds to the mean-square distance of each electron from the center of mass.

\begin{figure}[t]
\includegraphics[width=0.48\textwidth]{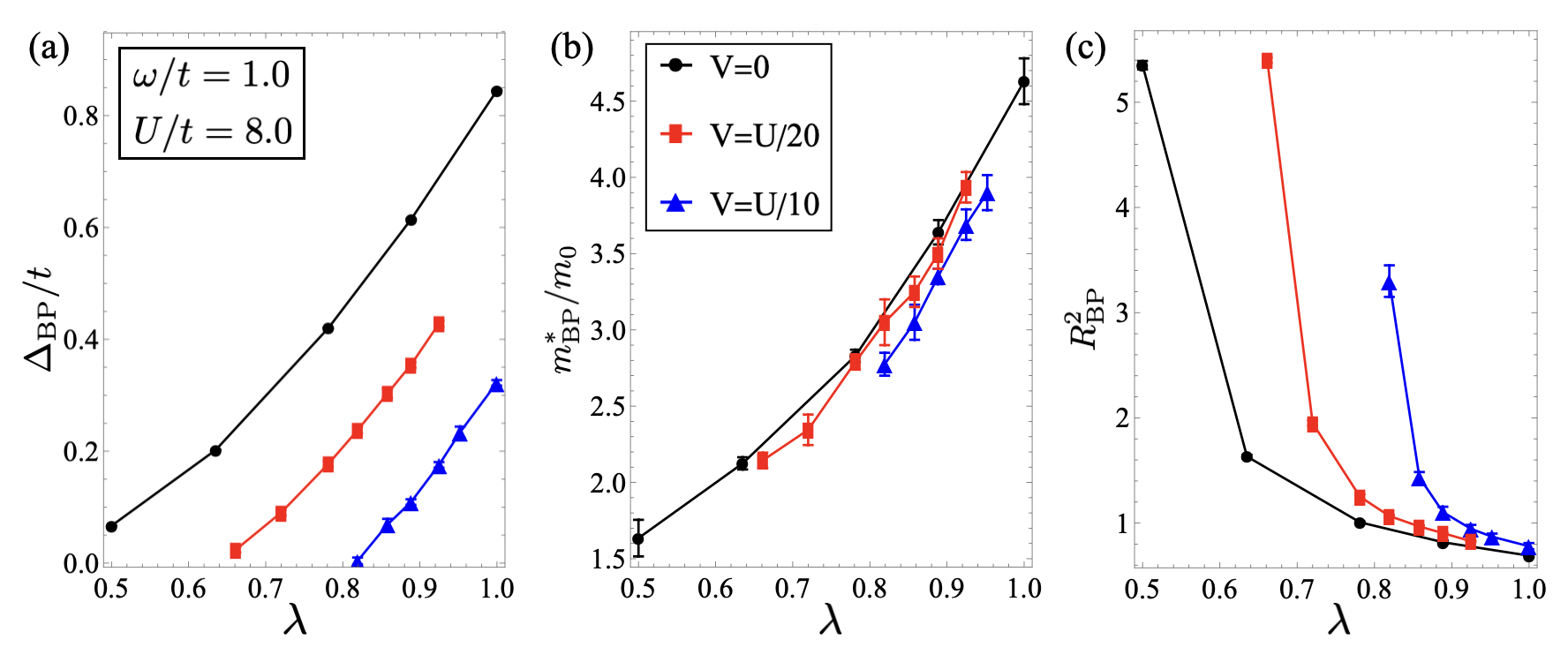}
\caption{Single-bipolaron properties computed from diagrammatic Monte Carlo simulations of Eq.~\eqref{Eq1} at $\omega/t=1.0$ as a function of the dimensionless coupling $\lambda$, for $U/t=8.0$ and Coulomb strengths $V=0$ (black dots), $V=U/20$ (red squares), and $V=U/10$ (blue triangles).
(a) Binding energy $\Delta_{\mathrm{BP}}$ in units of $t$.
(b) Effective mass $m^{*}_{\mathrm{BP}}/m_0$.
(c) Mean-square radius $R_{\mathrm{BP}}^2$.
Error bars denote one standard deviation (statistical).
}
\label{FIG2}
\end{figure}

\section{Superfluid transition temperature of bipolarons}
\label{sec:sec4}

In the dilute-carrier regime, sufficiently strong electron--phonon coupling can bind two electrons into a bipolaron. 
At low bipolaron density, these composite pairs may be viewed as an interacting 2D Bose gas whose phase coherence is destroyed via a BKT transition. 
A commonly used estimate relates the transition temperature to the superfluid stiffness, and for a dilute bosonic gas it takes the form
$T_c \simeq 1.84\, \rho_{\rm BP}/m_{\rm BP}^*$~\cite{PhysRevB.37.4936,PhysRevLett.87.270402,PhysRevLett.100.140405}, with where $\rho_{\rm BP}$ is the bipolaron density and $m_{\rm BP}^{*}$ is the single-bipolaron effective mass. In this work we do not simulate a finite-density bipolaron gas explicitly. 
Instead, we construct a dilute-limit BKT estimate from the single-bipolaron effective mass and size obtained in Sec.~\ref{sec:sec3}.

To approximately account for the additional suppression of the BKT transition temperature by long-range Coulomb repulsion in a 2D Bose gas, we introduce a multiplicative prefactor $C$.
Motivated by the estimates reported in Ref.~\cite{PhysRevLett.130.236001} for a 2D bosonic gas with Coluomb interactions, we use $C=0.85$ as a representative value when quoting Coulomb-reduced transition temperatures.

With this convention, we estimate
\begin{equation}
T_c \approx C \times 1.84 \times \frac{\rho_{\rm BP}}{m_{\rm BP}^{*}}.
\label{EqTc_rho}
\end{equation}
The estimate is meaningful in the non-overlapping regime, where the typical inter-bipolaron spacing is larger than the bipolaron size.

To obtain an optimized dilute-limit estimate using only single-bipolaron inputs, we choose $\rho_{\rm BP}$ to maximize $T_c$ subject to the non-overlap condition.
We take the characteristic spacing to be of order the bipolaron radius $R_{\rm BP}$ (in lattice units), which yields the estimate
$\rho_{\rm BP}\sim \min\!\left[1/(\pi R_{\rm BP}^2),\,1/\pi\right]$.
Substituting this into Eq.~\eqref{EqTc_rho} leads to
\begin{equation}
T_c \approx
  \begin{cases}
C \cdot \dfrac{0.5}{m_{\rm BP}^{*}\, R_{\rm BP}^{2}}, & R_{\rm BP}^{2} \ge 1,\\[6pt]
C \cdot \dfrac{0.5}{m_{\rm BP}^{*}}, & R_{\rm BP}^{2} < 1,
  \end{cases}
\label{Eq2}
\end{equation}
where the numerical prefactor $0.5$ is adopted to match the convention used in Ref.~\cite{PhysRevX.13.011010}.
When quoting Coulomb-reduced transition temperatures we take $C=0.85$ for $V\neq 0$, while for $V=0$ we set $C=1$.
In the following, we denote this optimized dilute-limit estimate by $T_c$, note that it is an estimator constructed from single-bipolaron inputs rather than an explicit finite-density transition calculation.

\begin{figure}[t]
\includegraphics[width=0.47\textwidth]{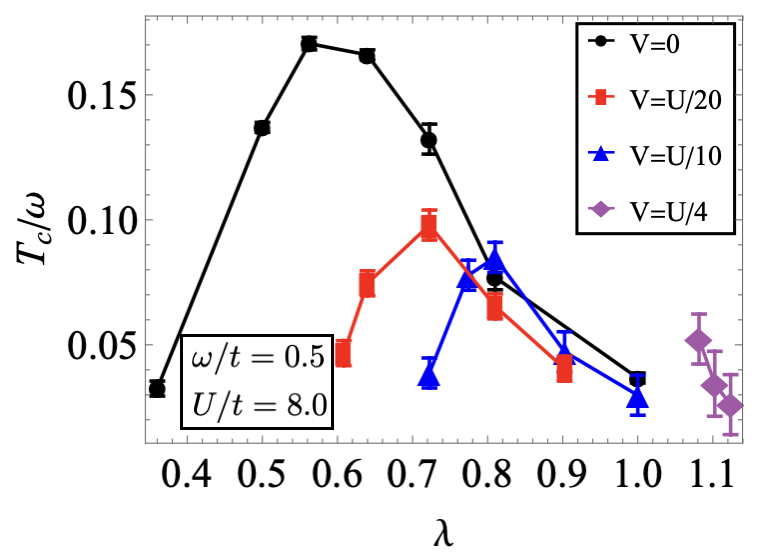}
\caption{Dilute-limit BKT estimates of $T_c/\omega$ for bond (SSH) bipolarons as a function of the dimensionless coupling $\lambda$, for Coulomb strengths $V=0$ (black dots), $V=U/20$ (red squares), $V=U/10$ (blue triangles), and $V=U/4$ (purple diamonds), at fixed $\omega/t=0.5$ and $U/t=8.0$.
The estimates are obtained from Eq.~\ref{Eq2} using diagrammatic Monte Carlo results for the single-bipolaron effective mass $m^{*}_{\mathrm{BP}}$ and mean-square radius $R^2_{\mathrm{BP}}$ in the two-electron sector.
Error bars denote one standard deviation (statistical).
}
\label{FIG3}
\end{figure}

\section{Results and discussion}
\label{sec:sec5}

Figure~\ref{FIG1} shows the optimized dilute-limit BKT estimates of $T_c/\omega$ for bond (SSH) bipolarons as a function of the dimensionless coupling $\lambda$ at $\omega/t=1.0$ and $U/t=8.0$, for Coulomb strengths $V=0$, $U/20$, and $U/10$.
In all cases the estimated $T_c$ exhibits a dome-like dependence on $\lambda$, consistent with the $V=0$ results reported previously and the general trends discussed in Ref.~\cite{PhysRevX.13.011010, PhysRevB.108.L220502}.
As the long-range repulsion increases, the maximal $T_c$ decreases and the optimal $\lambda$ window shifts; the suppression of the peak becomes stronger when $V$ is increased from $U/20$ to $U/10$.
To provide a conventional reference scale, we compare our optimized dilute-limit BKT estimates with McMillan's phenomenological expression for phonon-mediated superconductors within the Migdal--Eliashberg framework~\cite{Eliashberg1960, McMillan1968, Carbotte1990},
\begin{equation}
\frac{T_c}{\omega}=\frac{1}{1.45}\exp\!\left[-1.04\,\frac{1+\lambda}{\lambda-\mu^{*}(1+0.62\lambda)}\right],
\end{equation}
where $\omega$ denotes a characteristic phonon-frequency scale, $\lambda$ is the (Eliashberg) electron--phonon coupling constant, and $\mu^{*}$ is the Coulomb pseudopotential.
For the commonly used value $\mu^*=0.12$, McMillan's formula gives $T_c/\omega \simeq 0.05$ at $\lambda=1$. We emphasize that McMillan's expression is not quantitatively applicable in the present bipolaronic regime, and that our $\lambda$ [Eq.~\ref{lambda}] does not correspond to the Eliashberg coupling constant. 
Accordingly, the McMillan value is used only as a broad order-of-magnitude reference for conventional adiabatic phonon-mediated superconductors. 
With this caveat, the optimized dilute-limit estimates reach peak values $T_c/\omega \sim 0.12$ in Fig~\ref{FIG1} over the relevant parameter window, remaining larger than the reference McMillan scale even in the presence of long-range repulsion.

To understand these trends, Fig.~\ref{FIG2} reports the underlying single-bipolaron properties at $\omega/t=1.0$ and $U/t=8.0$.
Figure~\ref{FIG2}(a) shows that increasing $V$ weakens bipolaron binding and shifts the onset of a stable bound state to larger $\lambda$, reflecting the competition between electron-phonon coupling and Coulomb repulsion.
Within the bound-state region, we further find that at fixed $\lambda$ the effective mass $m_{\mathrm{BP}}^{*}$ decreases as $V$ is increased [Fig.~\ref{FIG2}(b)], while the mean-square radius $R_{\mathrm{BP}}^{2}$ increases [Fig.~\ref{FIG2}(c)].
The increase of $R_{\mathrm{BP}}^{2}$ is expected, as long-range repulsion disfavors compact configurations and expands the pair in real space.
The concomitant reduction of $m_{\mathrm{BP}}^{*}$ in parts of the parameter range indicates that the dispersion can become less renormalized despite the expanded bound state.
Since our optimized dilute-limit estimate is controlled by the combined mass--size dependence entering Eq.~\eqref{Eq2} (schematically, $T_c\sim 1/[m_{\mathrm{BP}}^{*}R_{\mathrm{BP}}^{2}]$ in the non-overlapping regime), these competing trends determine the net suppression and the location of the maximal transition temperature in Fig.~\ref{FIG1}.

\begin{figure}[t]
\includegraphics[width=0.48\textwidth]{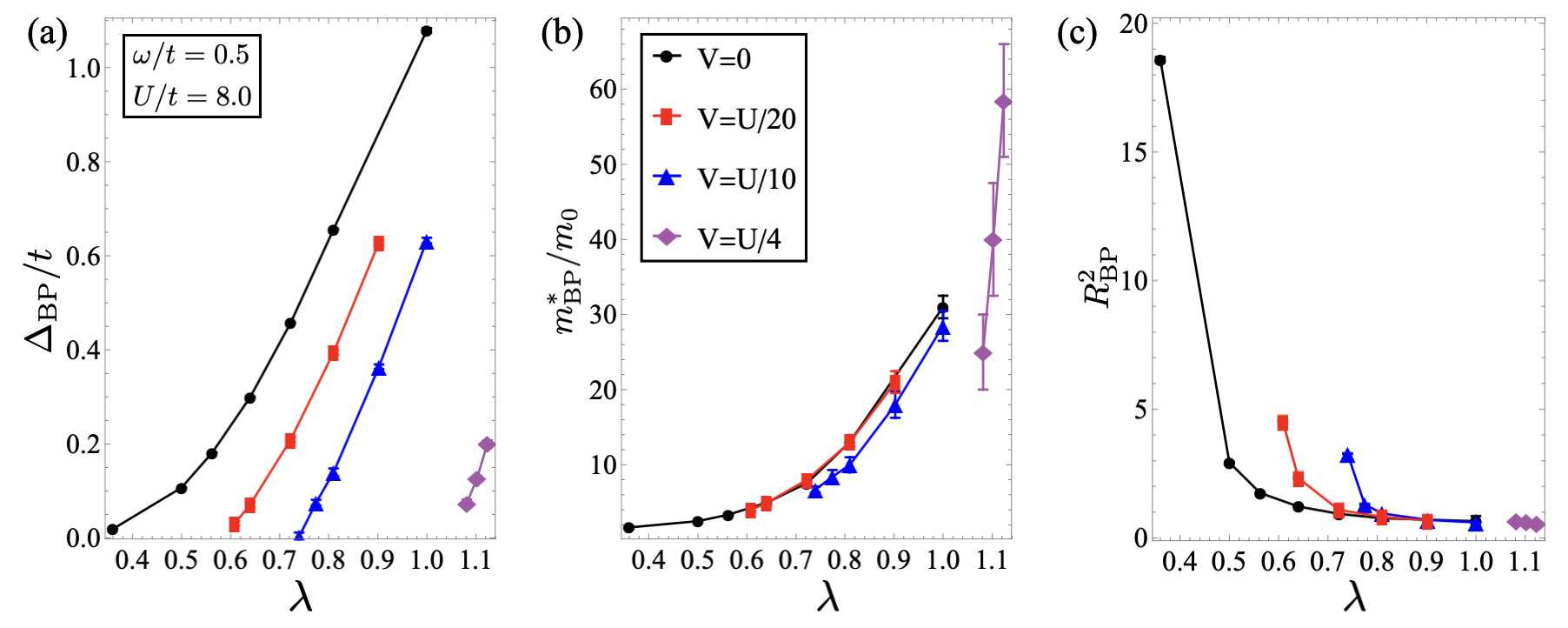}
\caption{Single-bipolaron properties from diagrammatic Monte Carlo simulations of Eq.~\eqref{Eq1} at $\omega/t=0.5$ as a function of the dimensionless coupling $\lambda$, for $U/t=8.0$ and Coulomb strengths $V=0$ (black dots), $V=U/20$ (red squares), $V=U/10$ (blue triangles), and $V=U/4$ (purple diamonds).
(a) Binding energy $\Delta_{\mathrm{BP}}$ in units of $t$.
(b) Effective mass $m^{*}_{\mathrm{BP}}/m_0$.
(c) Mean-square radius $R^2_{\mathrm{BP}}$.
Error bars denote one standard deviation (statistical).
}
\label{FIG4}
\end{figure}

We next turn to the more adiabatic regime $\omega/t=0.5$. In this regime, Coulomb repulsion has a more pronounced impact on the optimized BKT transition temperature, as shown in Fig.~\ref{FIG3}.
Compared to $\omega/t=1.0$, the dome in $T_c/\omega$ is generally shifted and more strongly suppressed at fixed $V/U$.
For $V\le U/10$, the optimized $T_c/\omega$ remains sizable over a broad window of moderate coupling, indicating that the SSH mechanism can still support relatively mobile composite pairs even in a deep adiabatic regime.

A particularly instructive case is the largest repulsion $V=U/4$, for which the $T_c$ dome is strongly suppressed and the optimized $T_c/\omega$ falls well below the reference scale $\sim 0.05$.
The microscopic origin of this suppression is evident from the single-bipolaron diagnostics in Fig.~\ref{FIG4}.
Figure~\ref{FIG4}(a) shows that strong long-range repulsion shifts the onset of a stable bipolaron to substantially larger $\lambda$, i.e., stronger SSH coupling is required to overcome the repulsion and form a bound state.
Moreover, once a bound state forms at large $\lambda$, Fig.~\ref{FIG4}(c) indicates that it becomes very compact, with $R_{\rm BP}^2$ approaching the lattice-scale minimum ($\sim 1$).
This compactification, however, is accompanied by a dramatic enhancement of the effective mass $m_{\rm BP}^*$ [Fig.~\ref{FIG4}(b)], reaching values an order of magnitude larger than for weaker repulsion.
Within our optimized dilute-limit estimate (Sec.~\ref{sec:sec4}), the BKT transition temperature is controlled parametrically by $T_c\propto 1/(m_{\rm BP}^* R_{\rm BP}^2)$ in the non-overlapping regime.
Consequently, for $V=U/4$ the heavy-mass penalty dominates over the reduced size, leading to a strongly suppressed $T_c$.
In particular, once $R_{\rm BP}^2\!\to\!1$ the non-overlap constraint effectively saturates the optimized density in Eq.~\eqref{Eq2}, so further compactification can no longer compensate the rapid growth of $m_{\rm BP}^*$; the increase of the effective mass therefore directly drives $T_c$ downward.
This behavior is consistent with a regime in which the pair remains tightly bound but its dispersion becomes extremely flat due to enhanced lattice dressing, so the bipolaron mobility (set by $m_{\rm BP}^*$) rather than its size becomes the limiting factor for the transition temperature.

\begin{figure}
\centering
\includegraphics[width=0.47\textwidth]{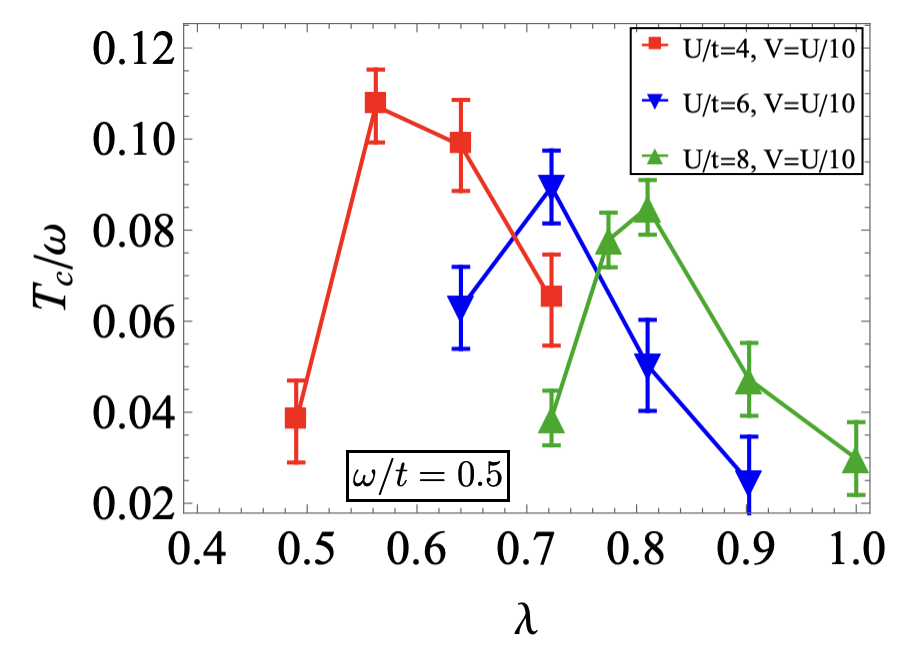}
\caption{Optimized dilute-limit BKT estimates of $T_c/\omega$ as a function of the dimensionless coupling $\lambda$ for fixed $V=U/10$ and $\omega/t=0.5$, shown for $U/t=4$ (red squares), $6$ (blue down-triangles), and $8$ (green up-triangles).
Error bars denote one standard deviation (statistical).
}\label{FIG5}
\end{figure}

We finally examine how the optimized transition temperature depends on the on-site Hubbard repulsion $U$ when long-range repulsion is present.
Figure~\ref{FIG5} shows $T_c/\omega$ at $\omega/t=0.5$ for fixed ratio $V=U/10$ and $U/t=4, 6, 8$.
Two robust trends emerge from this figure.
First, the location of the optimal dome shifts to larger $\lambda$ as $U/t$ increases: the maximum moves from $\lambda\simeq 0.55$--$0.6$ for $U/t=4$ to $\lambda\simeq 0.7$--$0.75$ for $U/t=6$ and to $\lambda\simeq 0.8$ for $U/t=8$.
Second, the peak value shows a mild overall reduction as $U/t$ increases, and the dome becomes progressively narrower on the strong-coupling side.

It is important to note that in Fig.~\ref{FIG5} we keep the ratio $V=U/10$ fixed, so increasing $U$ simultaneously increases the absolute long-range repulsion scale $V$.
The resulting suppression of the peak $T_c$ at larger $U/t$ therefore reflects a combined effect of stronger local repulsion and stronger long-range repulsion, which together reshape the bipolaron effective mass and size entering Eq.~\eqref{Eq2}.
Within these values of $U$, the data suggest that the monotonic enhancement with $U<8$ reported for $V=0$~\cite{PhysRevX.13.011010} does not directly carry over to $V>0$.
A more systematic study at fixed absolute $V$ would be useful to disentangle these effects and is left for future work.

\section{Conclusion}
\label{sec:sec6}

In the dilute-carrier regime, sufficiently strong electron-phonon coupling can bind two electrons into a bipolaron. At low bipolaron density, these composite pairs may be viewed as an interacting 2D Bose gas whose phase coherence is destroyed via a BKT transition. In this work, however, we do not simulate a finite-density bipolaron gas explicitly. Instead, we use the numerically exact single-bipolaron effective mass and spatial extent obtained in the two-electron sector to construct an optimized dilute-limit estimate of the transition temperature. This procedure should therefore be viewed as a controlled single-bipolaron benchmark combined with a physically motivated estimator for the dilute regime, rather than as a parameter-free determination of the true finite-density $T_c$.

We find that long-range repulsion generally suppresses the optimized BKT transition temperature by reshaping both the stability window of the bound state and the mobility of the composite pair.
Nevertheless, $T_c$ remains sizable over a broad parameter range, including $\omega/t=1.0$ and $0.5$ at $V=U/10$, where the SSH mechanism still supports relatively light and compact bipolarons.
In the more adiabatic regime, strong repulsion can drive a pronounced mass enhancement (Fig.~\ref{FIG4}), which becomes the primary mechanism suppressing the optimized $T_c$.
Finally, for fixed ratio $V=U/10$, our results in $U$ (Fig.~\ref{FIG5}) suggests that the peak $T_c$ does not increase with $U$ for $U<8$ once long-range repulsion is present; disentangling the roles of local and long-range interactions would benefit from a more systematic study at fixed absolute $V$.

Overall, our results provide a quantitative, controlled baseline for assessing the robustness of SSH bipolaron-based superfluidity against long-range repulsion in two dimensions, and for guiding future extensions that incorporate finite-density effects and screening explicitly.

\begin{acknowledgments}
This work is supported by the National Natural Science Foundation of China (NSFC) under Grants No. 12204173 and 12275002, and the University Annual Scientific Research Plan of Anhui Province under Grant No 2022AH010013.

\end{acknowledgments}

\bibliography{dispersive}

\begin{thebibliography}{38}%
\makeatletter
\providecommand \@ifxundefined [1]{%
 \@ifx{#1\undefined}
}%
\providecommand \@ifnum [1]{%
 \ifnum #1\expandafter \@firstoftwo
 \else \expandafter \@secondoftwo
 \fi
}%
\providecommand \@ifx [1]{%
 \ifx #1\expandafter \@firstoftwo
 \else \expandafter \@secondoftwo
 \fi
}%
\providecommand \natexlab [1]{#1}%
\providecommand \enquote  [1]{``#1''}%
\providecommand \bibnamefont  [1]{#1}%
\providecommand \bibfnamefont [1]{#1}%
\providecommand \citenamefont [1]{#1}%
\providecommand \href@noop [0]{\@secondoftwo}%
\providecommand \href [0]{\begingroup \@sanitize@url \@href}%
\providecommand \@href[1]{\@@startlink{#1}\@@href}%
\providecommand \@@href[1]{\endgroup#1\@@endlink}%
\providecommand \@sanitize@url [0]{\catcode `\\12\catcode `\$12\catcode
  `\&12\catcode `\#12\catcode `\^12\catcode `\_12\catcode `\%12\relax}%
\providecommand \@@startlink[1]{}%
\providecommand \@@endlink[0]{}%
\providecommand \url  [0]{\begingroup\@sanitize@url \@url }%
\providecommand \@url [1]{\endgroup\@href {#1}{\urlprefix }}%
\providecommand \urlprefix  [0]{URL }%
\providecommand \Eprint [0]{\href }%
\providecommand \doibase [0]{http://dx.doi.org/}%
\providecommand \selectlanguage [0]{\@gobble}%
\providecommand \bibinfo  [0]{\@secondoftwo}%
\providecommand \bibfield  [0]{\@secondoftwo}%
\providecommand \translation [1]{[#1]}%
\providecommand \BibitemOpen [0]{}%
\providecommand \bibitemStop [0]{}%
\providecommand \bibitemNoStop [0]{.\EOS\space}%
\providecommand \EOS [0]{\spacefactor3000\relax}%
\providecommand \BibitemShut  [1]{\csname bibitem#1\endcsname}%
\let\auto@bib@innerbib\@empty
\bibitem [{\citenamefont {Landau}(1933)}]{Landau33}%
  \BibitemOpen
  \bibfield  {author} {\bibinfo {author} {\bibfnamefont {L.~D.}\ \bibnamefont
  {Landau}},\ }\bibfield  {title} {\enquote {\bibinfo {title} {The movement of
  electrons in the crystal lattice},}\ }\href@noop {} {\bibfield  {journal}
  {\bibinfo  {journal} {Physikalische Zeitschrift der Sowjetunion}\ }\textbf
  {\bibinfo {volume} {3}},\ \bibinfo {pages} {664} (\bibinfo {year}
  {1933})}\BibitemShut {NoStop}%
\bibitem [{\citenamefont {H.~Fröhlich}\ and\ \citenamefont
  {Zienau}(1950)}]{Frohlich50}%
  \BibitemOpen
  \bibfield  {author} {\bibinfo {author} {\bibfnamefont {H.~Pelzer}\
  \bibnamefont {H.~Fröhlich}}\ and\ \bibinfo {author} {\bibfnamefont
  {S.}~\bibnamefont {Zienau}},\ }\bibfield  {title} {\enquote {\bibinfo {title}
  {Xx. properties of slow electrons in polar materials},}\ }\href {\doibase
  10.1080/14786445008521794} {\bibfield  {journal} {\bibinfo  {journal} {The
  London, Edinburgh, and Dublin Philosophical Magazine and Journal of Science}\
  }\textbf {\bibinfo {volume} {41}},\ \bibinfo {pages} {221--242} (\bibinfo
  {year} {1950})}\BibitemShut {NoStop}%
\bibitem [{\citenamefont {Feynman}(1955)}]{Feynman:1955du}%
  \BibitemOpen
  \bibfield  {author} {\bibinfo {author} {\bibfnamefont {R.~P.}\ \bibnamefont
  {Feynman}},\ }\bibfield  {title} {\enquote {\bibinfo {title} {Electron motion
  in crystal lattices},}\ }\href {\doibase 10.1103/PhysRev.97.660} {\bibfield
  {journal} {\bibinfo  {journal} {Phys. Rev.}\ }\textbf {\bibinfo {volume}
  {97}},\ \bibinfo {pages} {660--665} (\bibinfo {year} {1955})}\BibitemShut
  {NoStop}%
\bibitem [{\citenamefont {Schultz}(1959)}]{Schultz:1959el}%
  \BibitemOpen
  \bibfield  {author} {\bibinfo {author} {\bibfnamefont {T.~D.}\ \bibnamefont
  {Schultz}},\ }\bibfield  {title} {\enquote {\bibinfo {title} {Slow electrons
  in polar crystals: Self-energy, mass, and mobility},}\ }\href {\doibase
  10.1103/PhysRev.116.526} {\bibfield  {journal} {\bibinfo  {journal} {Phys.
  Rev.}\ }\textbf {\bibinfo {volume} {116}},\ \bibinfo {pages} {526--543}
  (\bibinfo {year} {1959})}\BibitemShut {NoStop}%
\bibitem [{\citenamefont {Holstein}(1959)}]{Holstein59}%
  \BibitemOpen
  \bibfield  {author} {\bibinfo {author} {\bibfnamefont {T}~\bibnamefont
  {Holstein}},\ }\bibfield  {title} {\enquote {\bibinfo {title} {Studies of
  polaron motion: Part i. the molecular-crystal model},}\ }\href {\doibase
  https://doi.org/10.1016/0003-4916(59)90002-8} {\bibfield  {journal} {\bibinfo
   {journal} {Annals of Physics}\ }\textbf {\bibinfo {volume} {8}},\ \bibinfo
  {pages} {325--342} (\bibinfo {year} {1959})}\BibitemShut {NoStop}%
\bibitem [{\citenamefont {Alexandrov}\ and\ \citenamefont
  {Kornilovitch}(1999)}]{Alexandrov:1999fy}%
  \BibitemOpen
  \bibfield  {author} {\bibinfo {author} {\bibfnamefont {A.~S.}\ \bibnamefont
  {Alexandrov}}\ and\ \bibinfo {author} {\bibfnamefont {P.~E.}\ \bibnamefont
  {Kornilovitch}},\ }\bibfield  {title} {\enquote {\bibinfo {title} {Mobile
  small polaron},}\ }\href {\doibase 10.1103/PhysRevLett.82.807} {\bibfield
  {journal} {\bibinfo  {journal} {Phys. Rev. Lett.}\ }\textbf {\bibinfo
  {volume} {82}},\ \bibinfo {pages} {807--810} (\bibinfo {year}
  {1999})}\BibitemShut {NoStop}%
\bibitem [{\citenamefont {Holstein}(2000)}]{Holstein2000}%
  \BibitemOpen
  \bibfield  {author} {\bibinfo {author} {\bibfnamefont {T.}~\bibnamefont
  {Holstein}},\ }\bibfield  {title} {\enquote {\bibinfo {title} {Studies of
  polaron motion: Part ii. the “small” polaron},}\ }\href {\doibase
  https://doi.org/10.1006/aphy.2000.6021} {\bibfield  {journal} {\bibinfo
  {journal} {Annals of Physics}\ }\textbf {\bibinfo {volume} {281}},\ \bibinfo
  {pages} {725--773} (\bibinfo {year} {2000})}\BibitemShut {NoStop}%
\bibitem [{\citenamefont {Ragni}\ \emph {et~al.}(2025)\citenamefont {Ragni},
  \citenamefont {Mi\ifmmode \check{s}\else \v{s}\fi{}ki\ifmmode~\acute{c}\else
  \'{c}\fi{}}, \citenamefont {Hahn}, \citenamefont {Prokof'ev}, \citenamefont
  {Bari\ifmmode \check{s}\else \v{s}\fi{}i\ifmmode~\acute{c}\else \'{c}\fi{}},
  \citenamefont {Nagaosa}, \citenamefont {Franchini},\ and\ \citenamefont
  {Mishchenko}}]{6127-phps}%
  \BibitemOpen
  \bibfield  {author} {\bibinfo {author} {\bibfnamefont {Stefano}\ \bibnamefont
  {Ragni}}, \bibinfo {author} {\bibfnamefont {Tomislav}\ \bibnamefont
  {Mi\ifmmode \check{s}\else \v{s}\fi{}ki\ifmmode~\acute{c}\else \'{c}\fi{}}},
  \bibinfo {author} {\bibfnamefont {Thomas}\ \bibnamefont {Hahn}}, \bibinfo
  {author} {\bibfnamefont {Nikolay}\ \bibnamefont {Prokof'ev}}, \bibinfo
  {author} {\bibfnamefont {Osor~S.}\ \bibnamefont {Bari\ifmmode \check{s}\else
  \v{s}\fi{}i\ifmmode~\acute{c}\else \'{c}\fi{}}}, \bibinfo {author}
  {\bibfnamefont {Naoto}\ \bibnamefont {Nagaosa}}, \bibinfo {author}
  {\bibfnamefont {Cesare}\ \bibnamefont {Franchini}}, \ and\ \bibinfo {author}
  {\bibfnamefont {Andrey~S.}\ \bibnamefont {Mishchenko}},\ }\bibfield  {title}
  {\enquote {\bibinfo {title} {Polarons with arbitrary nonlinear
  electron-phonon interaction},}\ }\href {\doibase 10.1103/6127-phps}
  {\bibfield  {journal} {\bibinfo  {journal} {Phys. Rev. Res.}\ }\textbf
  {\bibinfo {volume} {7}},\ \bibinfo {pages} {043304} (\bibinfo {year}
  {2025})}\BibitemShut {NoStop}%
\bibitem [{\citenamefont {Zhang}\ \emph
  {et~al.}(2023{\natexlab{a}})\citenamefont {Zhang}, \citenamefont {Kuklov},
  \citenamefont {Prokof'ev},\ and\ \citenamefont
  {Svistunov}}]{PhysRevB.108.245127}%
  \BibitemOpen
  \bibfield  {author} {\bibinfo {author} {\bibfnamefont {Zhongjin}\
  \bibnamefont {Zhang}}, \bibinfo {author} {\bibfnamefont {Anatoly}\
  \bibnamefont {Kuklov}}, \bibinfo {author} {\bibfnamefont {Nikolay}\
  \bibnamefont {Prokof'ev}}, \ and\ \bibinfo {author} {\bibfnamefont {Boris}\
  \bibnamefont {Svistunov}},\ }\bibfield  {title} {\enquote {\bibinfo {title}
  {Soliton states from quadratic electron-phonon interaction},}\ }\href
  {\doibase 10.1103/PhysRevB.108.245127} {\bibfield  {journal} {\bibinfo
  {journal} {Phys. Rev. B}\ }\textbf {\bibinfo {volume} {108}},\ \bibinfo
  {pages} {245127} (\bibinfo {year} {2023}{\natexlab{a}})}\BibitemShut
  {NoStop}%
\bibitem [{\citenamefont {Ragni}\ \emph {et~al.}(2023)\citenamefont {Ragni},
  \citenamefont {Hahn}, \citenamefont {Zhang}, \citenamefont {Prokof'ev},
  \citenamefont {Kuklov}, \citenamefont {Klimin}, \citenamefont {Houtput},
  \citenamefont {Svistunov}, \citenamefont {Tempere}, \citenamefont {Nagaosa},
  \citenamefont {Franchini},\ and\ \citenamefont
  {Mishchenko}}]{PhysRevB.107.L121109}%
  \BibitemOpen
  \bibfield  {author} {\bibinfo {author} {\bibfnamefont {Stefano}\ \bibnamefont
  {Ragni}}, \bibinfo {author} {\bibfnamefont {Thomas}\ \bibnamefont {Hahn}},
  \bibinfo {author} {\bibfnamefont {Zhongjin}\ \bibnamefont {Zhang}}, \bibinfo
  {author} {\bibfnamefont {Nikolay}\ \bibnamefont {Prokof'ev}}, \bibinfo
  {author} {\bibfnamefont {Anatoly}\ \bibnamefont {Kuklov}}, \bibinfo {author}
  {\bibfnamefont {Serghei}\ \bibnamefont {Klimin}}, \bibinfo {author}
  {\bibfnamefont {Matthew}\ \bibnamefont {Houtput}}, \bibinfo {author}
  {\bibfnamefont {Boris}\ \bibnamefont {Svistunov}}, \bibinfo {author}
  {\bibfnamefont {Jacques}\ \bibnamefont {Tempere}}, \bibinfo {author}
  {\bibfnamefont {Naoto}\ \bibnamefont {Nagaosa}}, \bibinfo {author}
  {\bibfnamefont {Cesare}\ \bibnamefont {Franchini}}, \ and\ \bibinfo {author}
  {\bibfnamefont {Andrey~S.}\ \bibnamefont {Mishchenko}},\ }\bibfield  {title}
  {\enquote {\bibinfo {title} {Polaron with quadratic electron-phonon
  interaction},}\ }\href {\doibase 10.1103/PhysRevB.107.L121109} {\bibfield
  {journal} {\bibinfo  {journal} {Phys. Rev. B}\ }\textbf {\bibinfo {volume}
  {107}},\ \bibinfo {pages} {L121109} (\bibinfo {year} {2023})}\BibitemShut
  {NoStop}%
\bibitem [{\citenamefont {Kornilovitch}\ and\ \citenamefont
  {Pike}(1997)}]{PhysRevB.55.R8634}%
  \BibitemOpen
  \bibfield  {author} {\bibinfo {author} {\bibfnamefont {P.~E.}\ \bibnamefont
  {Kornilovitch}}\ and\ \bibinfo {author} {\bibfnamefont {E.~R.}\ \bibnamefont
  {Pike}},\ }\bibfield  {title} {\enquote {\bibinfo {title} {Polaron effective
  mass from monte carlo simulations},}\ }\href {\doibase
  10.1103/PhysRevB.55.R8634} {\bibfield  {journal} {\bibinfo  {journal} {Phys.
  Rev. B}\ }\textbf {\bibinfo {volume} {55}},\ \bibinfo {pages} {R8634--R8637}
  (\bibinfo {year} {1997})}\BibitemShut {NoStop}%
\bibitem [{\citenamefont {Bon\ifmmode~\check{c}\else \v{c}\fi{}a}\ \emph
  {et~al.}(2000)\citenamefont {Bon\ifmmode~\check{c}\else \v{c}\fi{}a},
  \citenamefont {Katra\ifmmode~\check{s}\else \v{s}\fi{}nik},\ and\
  \citenamefont {Trugman}}]{PhysRevLett.84.3153}%
  \BibitemOpen
  \bibfield  {author} {\bibinfo {author} {\bibfnamefont {J.}~\bibnamefont
  {Bon\ifmmode~\check{c}\else \v{c}\fi{}a}}, \bibinfo {author} {\bibfnamefont
  {T.}~\bibnamefont {Katra\ifmmode~\check{s}\else \v{s}\fi{}nik}}, \ and\
  \bibinfo {author} {\bibfnamefont {S.~A.}\ \bibnamefont {Trugman}},\
  }\bibfield  {title} {\enquote {\bibinfo {title} {Mobile bipolaron},}\ }\href
  {\doibase 10.1103/PhysRevLett.84.3153} {\bibfield  {journal} {\bibinfo
  {journal} {Phys. Rev. Lett.}\ }\textbf {\bibinfo {volume} {84}},\ \bibinfo
  {pages} {3153--3156} (\bibinfo {year} {2000})}\BibitemShut {NoStop}%
\bibitem [{\citenamefont {Macridin}\ \emph {et~al.}(2004)\citenamefont
  {Macridin}, \citenamefont {Sawatzky},\ and\ \citenamefont
  {Jarrell}}]{PhysRevB.69.245111}%
  \BibitemOpen
  \bibfield  {author} {\bibinfo {author} {\bibfnamefont {A.}~\bibnamefont
  {Macridin}}, \bibinfo {author} {\bibfnamefont {G.~A.}\ \bibnamefont
  {Sawatzky}}, \ and\ \bibinfo {author} {\bibfnamefont {M.}~\bibnamefont
  {Jarrell}},\ }\bibfield  {title} {\enquote {\bibinfo {title} {Two-dimensional
  hubbard-holstein bipolaron},}\ }\href {\doibase 10.1103/PhysRevB.69.245111}
  {\bibfield  {journal} {\bibinfo  {journal} {Phys. Rev. B}\ }\textbf {\bibinfo
  {volume} {69}},\ \bibinfo {pages} {245111} (\bibinfo {year}
  {2004})}\BibitemShut {NoStop}%
\bibitem [{\citenamefont {Marchand}\ \emph {et~al.}(2010)\citenamefont
  {Marchand}, \citenamefont {De~Filippis}, \citenamefont {Cataudella},
  \citenamefont {Berciu}, \citenamefont {Nagaosa}, \citenamefont {Prokof'ev},
  \citenamefont {Mishchenko},\ and\ \citenamefont
  {Stamp}}]{PhysRevLett.105.266605}%
  \BibitemOpen
  \bibfield  {author} {\bibinfo {author} {\bibfnamefont {D.~J.~J.}\
  \bibnamefont {Marchand}}, \bibinfo {author} {\bibfnamefont {G.}~\bibnamefont
  {De~Filippis}}, \bibinfo {author} {\bibfnamefont {V.}~\bibnamefont
  {Cataudella}}, \bibinfo {author} {\bibfnamefont {M.}~\bibnamefont {Berciu}},
  \bibinfo {author} {\bibfnamefont {N.}~\bibnamefont {Nagaosa}}, \bibinfo
  {author} {\bibfnamefont {N.~V.}\ \bibnamefont {Prokof'ev}}, \bibinfo {author}
  {\bibfnamefont {A.~S.}\ \bibnamefont {Mishchenko}}, \ and\ \bibinfo {author}
  {\bibfnamefont {P.~C.~E.}\ \bibnamefont {Stamp}},\ }\bibfield  {title}
  {\enquote {\bibinfo {title} {Sharp transition for single polarons in the
  one-dimensional su-schrieffer-heeger model},}\ }\href {\doibase
  10.1103/PhysRevLett.105.266605} {\bibfield  {journal} {\bibinfo  {journal}
  {Phys. Rev. Lett.}\ }\textbf {\bibinfo {volume} {105}},\ \bibinfo {pages}
  {266605} (\bibinfo {year} {2010})}\BibitemShut {NoStop}%
\bibitem [{\citenamefont {Sous}\ \emph {et~al.}(2018)\citenamefont {Sous},
  \citenamefont {Chakraborty}, \citenamefont {Krems},\ and\ \citenamefont
  {Berciu}}]{PhysRevLett.121.247001}%
  \BibitemOpen
  \bibfield  {author} {\bibinfo {author} {\bibfnamefont {John}\ \bibnamefont
  {Sous}}, \bibinfo {author} {\bibfnamefont {Monodeep}\ \bibnamefont
  {Chakraborty}}, \bibinfo {author} {\bibfnamefont {Roman~V.}\ \bibnamefont
  {Krems}}, \ and\ \bibinfo {author} {\bibfnamefont {Mona}\ \bibnamefont
  {Berciu}},\ }\bibfield  {title} {\enquote {\bibinfo {title} {Light bipolarons
  stabilized by peierls electron-phonon coupling},}\ }\href {\doibase
  10.1103/PhysRevLett.121.247001} {\bibfield  {journal} {\bibinfo  {journal}
  {Phys. Rev. Lett.}\ }\textbf {\bibinfo {volume} {121}},\ \bibinfo {pages}
  {247001} (\bibinfo {year} {2018})}\BibitemShut {NoStop}%
\bibitem [{\citenamefont {Kim}\ \emph {et~al.}(2024)\citenamefont {Kim},
  \citenamefont {Han},\ and\ \citenamefont {Sous}}]{PhysRevB.109.L220502}%
  \BibitemOpen
  \bibfield  {author} {\bibinfo {author} {\bibfnamefont {Kyung-Su}\
  \bibnamefont {Kim}}, \bibinfo {author} {\bibfnamefont {Zhaoyu}\ \bibnamefont
  {Han}}, \ and\ \bibinfo {author} {\bibfnamefont {John}\ \bibnamefont
  {Sous}},\ }\bibfield  {title} {\enquote {\bibinfo {title} {Semiclassical
  theory of bipolaronic superconductivity in a bond-modulated electron-phonon
  model},}\ }\href {\doibase 10.1103/PhysRevB.109.L220502} {\bibfield
  {journal} {\bibinfo  {journal} {Phys. Rev. B}\ }\textbf {\bibinfo {volume}
  {109}},\ \bibinfo {pages} {L220502} (\bibinfo {year} {2024})}\BibitemShut
  {NoStop}%
\bibitem [{\citenamefont {Zhang}\ \emph {et~al.}(2021)\citenamefont {Zhang},
  \citenamefont {Prokof'ev},\ and\ \citenamefont
  {Svistunov}}]{PhysRevB.104.035143}%
  \BibitemOpen
  \bibfield  {author} {\bibinfo {author} {\bibfnamefont {C.}~\bibnamefont
  {Zhang}}, \bibinfo {author} {\bibfnamefont {N.~V.}\ \bibnamefont
  {Prokof'ev}}, \ and\ \bibinfo {author} {\bibfnamefont {B.~V.}\ \bibnamefont
  {Svistunov}},\ }\bibfield  {title} {\enquote {\bibinfo {title}
  {Peierls/su-schrieffer-heeger polarons in two dimensions},}\ }\href {\doibase
  10.1103/PhysRevB.104.035143} {\bibfield  {journal} {\bibinfo  {journal}
  {Phys. Rev. B}\ }\textbf {\bibinfo {volume} {104}},\ \bibinfo {pages}
  {035143} (\bibinfo {year} {2021})}\BibitemShut {NoStop}%
\bibitem [{\citenamefont {Zhang}(2025{\natexlab{a}})}]{ckbn-jp9t}%
  \BibitemOpen
  \bibfield  {author} {\bibinfo {author} {\bibfnamefont {Chao}\ \bibnamefont
  {Zhang}},\ }\bibfield  {title} {\enquote {\bibinfo {title} {Comprehensive
  study of bond bipolaron superconductivity on the triangular lattice},}\
  }\href {\doibase 10.1103/ckbn-jp9t} {\bibfield  {journal} {\bibinfo
  {journal} {Phys. Rev. B}\ }\textbf {\bibinfo {volume} {112}},\ \bibinfo
  {pages} {174520} (\bibinfo {year} {2025}{\natexlab{a}})}\BibitemShut
  {NoStop}%
\bibitem [{\citenamefont {Carbone}\ \emph {et~al.}(2021)\citenamefont
  {Carbone}, \citenamefont {Millis}, \citenamefont {Reichman},\ and\
  \citenamefont {Sous}}]{PhysRevB.104.L140307}%
  \BibitemOpen
  \bibfield  {author} {\bibinfo {author} {\bibfnamefont {M.~R.}\ \bibnamefont
  {Carbone}}, \bibinfo {author} {\bibfnamefont {A.~J.}\ \bibnamefont {Millis}},
  \bibinfo {author} {\bibfnamefont {D.~R.}\ \bibnamefont {Reichman}}, \ and\
  \bibinfo {author} {\bibfnamefont {J.}~\bibnamefont {Sous}},\ }\bibfield
  {title} {\enquote {\bibinfo {title} {Bond-peierls polaron: Moderate mass
  enhancement and current-carrying ground state},}\ }\href {\doibase
  10.1103/PhysRevB.104.L140307} {\bibfield  {journal} {\bibinfo  {journal}
  {Phys. Rev. B}\ }\textbf {\bibinfo {volume} {104}},\ \bibinfo {pages}
  {L140307} (\bibinfo {year} {2021})}\BibitemShut {NoStop}%
\bibitem [{\citenamefont {Zhang}(2023)}]{PhysRevB.108.075156}%
  \BibitemOpen
  \bibfield  {author} {\bibinfo {author} {\bibfnamefont {C.}~\bibnamefont
  {Zhang}},\ }\bibfield  {title} {\enquote {\bibinfo {title} {Effect of
  dispersive optical phonons on the properties of the bond su-schrieffer-heeger
  polaron},}\ }\href {\doibase 10.1103/PhysRevB.108.075156} {\bibfield
  {journal} {\bibinfo  {journal} {Phys. Rev. B}\ }\textbf {\bibinfo {volume}
  {108}},\ \bibinfo {pages} {075156} (\bibinfo {year} {2023})}\BibitemShut
  {NoStop}%
\bibitem [{\citenamefont {Zhang}(2024)}]{PhysRevB.109.165119}%
  \BibitemOpen
  \bibfield  {author} {\bibinfo {author} {\bibfnamefont {C.}~\bibnamefont
  {Zhang}},\ }\bibfield  {title} {\enquote {\bibinfo {title} {Light polarons
  with electron-phonon coupling},}\ }\href {\doibase
  10.1103/PhysRevB.109.165119} {\bibfield  {journal} {\bibinfo  {journal}
  {Phys. Rev. B}\ }\textbf {\bibinfo {volume} {109}},\ \bibinfo {pages}
  {165119} (\bibinfo {year} {2024})}\BibitemShut {NoStop}%
\bibitem [{\citenamefont {Zhang}\ \emph
  {et~al.}(2025{\natexlab{a}})\citenamefont {Zhang}, \citenamefont
  {Prokof'ev},\ and\ \citenamefont {Svistunov}}]{PhysRevB.111.184513}%
  \BibitemOpen
  \bibfield  {author} {\bibinfo {author} {\bibfnamefont {Chao}\ \bibnamefont
  {Zhang}}, \bibinfo {author} {\bibfnamefont {Nikolay}\ \bibnamefont
  {Prokof'ev}}, \ and\ \bibinfo {author} {\bibfnamefont {Boris}\ \bibnamefont
  {Svistunov}},\ }\bibfield  {title} {\enquote {\bibinfo {title} {Effects of
  phonon dispersion on bond-bipolaron superconductivity},}\ }\href {\doibase
  10.1103/PhysRevB.111.184513} {\bibfield  {journal} {\bibinfo  {journal}
  {Phys. Rev. B}\ }\textbf {\bibinfo {volume} {111}},\ \bibinfo {pages}
  {184513} (\bibinfo {year} {2025}{\natexlab{a}})}\BibitemShut {NoStop}%
\bibitem [{\citenamefont {Zhang}(2025{\natexlab{b}})}]{arXivchao2025}%
  \BibitemOpen
  \bibfield  {author} {\bibinfo {author} {\bibfnamefont {Chao}\ \bibnamefont
  {Zhang}},\ }\bibfield  {title} {\enquote {\bibinfo {title} {Robustness of
  bipolaronic superconductivity to electron-density-phonon coupling},}\ }\href
  {https://arxiv.org/pdf/2511.06350} {\bibfield  {journal} {\bibinfo  {journal}
  {arXiv:2511.06350}\ } (\bibinfo {year} {2025}{\natexlab{b}})}\BibitemShut
  {NoStop}%
\bibitem [{\citenamefont {Zhang}\ \emph
  {et~al.}(2025{\natexlab{b}})\citenamefont {Zhang}, \citenamefont {Kuklov},
  \citenamefont {Prokof'ev},\ and\ \citenamefont
  {Svistunov}}]{PhysRevB.111.134504}%
  \BibitemOpen
  \bibfield  {author} {\bibinfo {author} {\bibfnamefont {Zhongjin}\
  \bibnamefont {Zhang}}, \bibinfo {author} {\bibfnamefont {Anatoly}\
  \bibnamefont {Kuklov}}, \bibinfo {author} {\bibfnamefont {Nikolay}\
  \bibnamefont {Prokof'ev}}, \ and\ \bibinfo {author} {\bibfnamefont {Boris}\
  \bibnamefont {Svistunov}},\ }\bibfield  {title} {\enquote {\bibinfo {title}
  {Superconductivity of bipolarons from quadratic electron-phonon
  interaction},}\ }\href {\doibase 10.1103/PhysRevB.111.134504} {\bibfield
  {journal} {\bibinfo  {journal} {Phys. Rev. B}\ }\textbf {\bibinfo {volume}
  {111}},\ \bibinfo {pages} {134504} (\bibinfo {year}
  {2025}{\natexlab{b}})}\BibitemShut {NoStop}%
\bibitem [{\citenamefont {Zhang}\ \emph
  {et~al.}(2023{\natexlab{b}})\citenamefont {Zhang}, \citenamefont {Sous},
  \citenamefont {Reichman}, \citenamefont {Berciu}, \citenamefont {Millis},
  \citenamefont {Prokof'ev},\ and\ \citenamefont
  {Svistunov}}]{PhysRevX.13.011010}%
  \BibitemOpen
  \bibfield  {author} {\bibinfo {author} {\bibfnamefont {C.}~\bibnamefont
  {Zhang}}, \bibinfo {author} {\bibfnamefont {J.}~\bibnamefont {Sous}},
  \bibinfo {author} {\bibfnamefont {D.~R.}\ \bibnamefont {Reichman}}, \bibinfo
  {author} {\bibfnamefont {M.}~\bibnamefont {Berciu}}, \bibinfo {author}
  {\bibfnamefont {A.~J.}\ \bibnamefont {Millis}}, \bibinfo {author}
  {\bibfnamefont {N.~V.}\ \bibnamefont {Prokof'ev}}, \ and\ \bibinfo {author}
  {\bibfnamefont {B.~V.}\ \bibnamefont {Svistunov}},\ }\bibfield  {title}
  {\enquote {\bibinfo {title} {Bipolaronic high-temperature
  superconductivity},}\ }\href {\doibase 10.1103/PhysRevX.13.011010} {\bibfield
   {journal} {\bibinfo  {journal} {Phys. Rev. X}\ }\textbf {\bibinfo {volume}
  {13}},\ \bibinfo {pages} {011010} (\bibinfo {year}
  {2023}{\natexlab{b}})}\BibitemShut {NoStop}%
\bibitem [{\citenamefont {Sous}\ \emph {et~al.}(2023)\citenamefont {Sous},
  \citenamefont {Zhang}, \citenamefont {Berciu}, \citenamefont {Reichman},
  \citenamefont {Svistunov}, \citenamefont {Prokof'ev},\ and\ \citenamefont
  {Millis}}]{PhysRevB.108.L220502}%
  \BibitemOpen
  \bibfield  {author} {\bibinfo {author} {\bibfnamefont {J.}~\bibnamefont
  {Sous}}, \bibinfo {author} {\bibfnamefont {C.}~\bibnamefont {Zhang}},
  \bibinfo {author} {\bibfnamefont {M.}~\bibnamefont {Berciu}}, \bibinfo
  {author} {\bibfnamefont {D.~R.}\ \bibnamefont {Reichman}}, \bibinfo {author}
  {\bibfnamefont {B.~V.}\ \bibnamefont {Svistunov}}, \bibinfo {author}
  {\bibfnamefont {N.~V.}\ \bibnamefont {Prokof'ev}}, \ and\ \bibinfo {author}
  {\bibfnamefont {A.~J.}\ \bibnamefont {Millis}},\ }\bibfield  {title}
  {\enquote {\bibinfo {title} {Bipolaronic superconductivity out of a coulomb
  gas},}\ }\href {\doibase 10.1103/PhysRevB.108.L220502} {\bibfield  {journal}
  {\bibinfo  {journal} {Phys. Rev. B}\ }\textbf {\bibinfo {volume} {108}},\
  \bibinfo {pages} {L220502} (\bibinfo {year} {2023})}\BibitemShut {NoStop}%
\bibitem [{\citenamefont {Zhang}\ \emph {et~al.}(2022)\citenamefont {Zhang},
  \citenamefont {Prokof'ev},\ and\ \citenamefont
  {Svistunov}}]{PhysRevB.105.L020501}%
  \BibitemOpen
  \bibfield  {author} {\bibinfo {author} {\bibfnamefont {C.}~\bibnamefont
  {Zhang}}, \bibinfo {author} {\bibfnamefont {N.~V.}\ \bibnamefont
  {Prokof'ev}}, \ and\ \bibinfo {author} {\bibfnamefont {B.~V.}\ \bibnamefont
  {Svistunov}},\ }\bibfield  {title} {\enquote {\bibinfo {title} {Bond
  bipolarons: Sign-free monte carlo approach},}\ }\href {\doibase
  10.1103/PhysRevB.105.L020501} {\bibfield  {journal} {\bibinfo  {journal}
  {Phys. Rev. B}\ }\textbf {\bibinfo {volume} {105}},\ \bibinfo {pages}
  {L020501} (\bibinfo {year} {2022})}\BibitemShut {NoStop}%
\bibitem [{\citenamefont {Su}\ \emph {et~al.}(1979)\citenamefont {Su},
  \citenamefont {Schrieffer},\ and\ \citenamefont
  {Heeger}}]{PhysRevLett.42.1698}%
  \BibitemOpen
  \bibfield  {author} {\bibinfo {author} {\bibfnamefont {W.~P.}\ \bibnamefont
  {Su}}, \bibinfo {author} {\bibfnamefont {J.~R.}\ \bibnamefont {Schrieffer}},
  \ and\ \bibinfo {author} {\bibfnamefont {A.~J.}\ \bibnamefont {Heeger}},\
  }\bibfield  {title} {\enquote {\bibinfo {title} {Solitons in
  polyacetylene},}\ }\href {\doibase 10.1103/PhysRevLett.42.1698} {\bibfield
  {journal} {\bibinfo  {journal} {Phys. Rev. Lett.}\ }\textbf {\bibinfo
  {volume} {42}},\ \bibinfo {pages} {1698--1701} (\bibinfo {year}
  {1979})}\BibitemShut {NoStop}%
\bibitem [{\citenamefont {Bari\ifmmode \check{s}\else
  \v{s}\fi{}i\ifmmode~\acute{c}\else \'{c}\fi{}}\ \emph
  {et~al.}(1970)\citenamefont {Bari\ifmmode \check{s}\else
  \v{s}\fi{}i\ifmmode~\acute{c}\else \'{c}\fi{}}, \citenamefont {Labb\'e},\
  and\ \citenamefont {Friedel}}]{PhysRevLett.25.919}%
  \BibitemOpen
  \bibfield  {author} {\bibinfo {author} {\bibfnamefont {S.}~\bibnamefont
  {Bari\ifmmode \check{s}\else \v{s}\fi{}i\ifmmode~\acute{c}\else \'{c}\fi{}}},
  \bibinfo {author} {\bibfnamefont {J.}~\bibnamefont {Labb\'e}}, \ and\
  \bibinfo {author} {\bibfnamefont {J.}~\bibnamefont {Friedel}},\ }\bibfield
  {title} {\enquote {\bibinfo {title} {Tight binding and transition-metal
  superconductivity},}\ }\href {\doibase 10.1103/PhysRevLett.25.919} {\bibfield
   {journal} {\bibinfo  {journal} {Phys. Rev. Lett.}\ }\textbf {\bibinfo
  {volume} {25}},\ \bibinfo {pages} {919--922} (\bibinfo {year}
  {1970})}\BibitemShut {NoStop}%
\bibitem [{\citenamefont {Bari\ifmmode \check{s}\else
  \v{s}\fi{}i\ifmmode~\acute{c}\else
  \'{c}\fi{}}(1972{\natexlab{a}})}]{PhysRevB.5.932}%
  \BibitemOpen
  \bibfield  {author} {\bibinfo {author} {\bibfnamefont {S.}~\bibnamefont
  {Bari\ifmmode \check{s}\else \v{s}\fi{}i\ifmmode~\acute{c}\else
  \'{c}\fi{}}},\ }\bibfield  {title} {\enquote {\bibinfo {title} {Rigid-atom
  electron-phonon coupling in the tight-binding approximation.i},}\ }\href
  {\doibase 10.1103/PhysRevB.5.932} {\bibfield  {journal} {\bibinfo  {journal}
  {Phys. Rev. B}\ }\textbf {\bibinfo {volume} {5}},\ \bibinfo {pages}
  {932--941} (\bibinfo {year} {1972}{\natexlab{a}})}\BibitemShut {NoStop}%
\bibitem [{\citenamefont {Bari\ifmmode \check{s}\else
  \v{s}\fi{}i\ifmmode~\acute{c}\else
  \'{c}\fi{}}(1972{\natexlab{b}})}]{PhysRevB.5.941}%
  \BibitemOpen
  \bibfield  {author} {\bibinfo {author} {\bibfnamefont {S.}~\bibnamefont
  {Bari\ifmmode \check{s}\else \v{s}\fi{}i\ifmmode~\acute{c}\else
  \'{c}\fi{}}},\ }\bibfield  {title} {\enquote {\bibinfo {title}
  {Self-consistent electron-phonon coupling in the tight-binding approximation.
  ii},}\ }\href {\doibase 10.1103/PhysRevB.5.941} {\bibfield  {journal}
  {\bibinfo  {journal} {Phys. Rev. B}\ }\textbf {\bibinfo {volume} {5}},\
  \bibinfo {pages} {941--951} (\bibinfo {year}
  {1972}{\natexlab{b}})}\BibitemShut {NoStop}%
\bibitem [{\citenamefont {Fisher}\ and\ \citenamefont
  {Hohenberg}(1988)}]{PhysRevB.37.4936}%
  \BibitemOpen
  \bibfield  {author} {\bibinfo {author} {\bibfnamefont {Daniel~S.}\
  \bibnamefont {Fisher}}\ and\ \bibinfo {author} {\bibfnamefont {P.~C.}\
  \bibnamefont {Hohenberg}},\ }\bibfield  {title} {\enquote {\bibinfo {title}
  {Dilute bose gas in two dimensions},}\ }\href {\doibase
  10.1103/PhysRevB.37.4936} {\bibfield  {journal} {\bibinfo  {journal} {Phys.
  Rev. B}\ }\textbf {\bibinfo {volume} {37}},\ \bibinfo {pages} {4936--4943}
  (\bibinfo {year} {1988})}\BibitemShut {NoStop}%
\bibitem [{\citenamefont {Prokof'ev}\ \emph {et~al.}(2001)\citenamefont
  {Prokof'ev}, \citenamefont {Ruebenacker},\ and\ \citenamefont
  {Svistunov}}]{PhysRevLett.87.270402}%
  \BibitemOpen
  \bibfield  {author} {\bibinfo {author} {\bibfnamefont {N.}~\bibnamefont
  {Prokof'ev}}, \bibinfo {author} {\bibfnamefont {O.}~\bibnamefont
  {Ruebenacker}}, \ and\ \bibinfo {author} {\bibfnamefont {B.}~\bibnamefont
  {Svistunov}},\ }\bibfield  {title} {\enquote {\bibinfo {title} {Critical
  point of a weakly interacting two-dimensional bose gas},}\ }\href {\doibase
  10.1103/PhysRevLett.87.270402} {\bibfield  {journal} {\bibinfo  {journal}
  {Phys. Rev. Lett.}\ }\textbf {\bibinfo {volume} {87}},\ \bibinfo {pages}
  {270402} (\bibinfo {year} {2001})}\BibitemShut {NoStop}%
\bibitem [{\citenamefont {Pilati}\ \emph {et~al.}(2008)\citenamefont {Pilati},
  \citenamefont {Giorgini},\ and\ \citenamefont
  {Prokof'ev}}]{PhysRevLett.100.140405}%
  \BibitemOpen
  \bibfield  {author} {\bibinfo {author} {\bibfnamefont {S.}~\bibnamefont
  {Pilati}}, \bibinfo {author} {\bibfnamefont {S.}~\bibnamefont {Giorgini}}, \
  and\ \bibinfo {author} {\bibfnamefont {N.}~\bibnamefont {Prokof'ev}},\
  }\bibfield  {title} {\enquote {\bibinfo {title} {Critical temperature of
  interacting bose gases in two and three dimensions},}\ }\href {\doibase
  10.1103/PhysRevLett.100.140405} {\bibfield  {journal} {\bibinfo  {journal}
  {Phys. Rev. Lett.}\ }\textbf {\bibinfo {volume} {100}},\ \bibinfo {pages}
  {140405} (\bibinfo {year} {2008})}\BibitemShut {NoStop}%
\bibitem [{\citenamefont {Zhang}\ \emph
  {et~al.}(2023{\natexlab{c}})\citenamefont {Zhang}, \citenamefont
  {Capogrosso-Sansone}, \citenamefont {Boninsegni}, \citenamefont {Prokof'ev},\
  and\ \citenamefont {Svistunov}}]{PhysRevLett.130.236001}%
  \BibitemOpen
  \bibfield  {author} {\bibinfo {author} {\bibfnamefont {C.}~\bibnamefont
  {Zhang}}, \bibinfo {author} {\bibfnamefont {B.}~\bibnamefont
  {Capogrosso-Sansone}}, \bibinfo {author} {\bibfnamefont {M.}~\bibnamefont
  {Boninsegni}}, \bibinfo {author} {\bibfnamefont {N.~V.}\ \bibnamefont
  {Prokof'ev}}, \ and\ \bibinfo {author} {\bibfnamefont {B.~V.}\ \bibnamefont
  {Svistunov}},\ }\bibfield  {title} {\enquote {\bibinfo {title}
  {Superconducting transition temperature of the bose one-component plasma},}\
  }\href {\doibase 10.1103/PhysRevLett.130.236001} {\bibfield  {journal}
  {\bibinfo  {journal} {Phys. Rev. Lett.}\ }\textbf {\bibinfo {volume} {130}},\
  \bibinfo {pages} {236001} (\bibinfo {year} {2023}{\natexlab{c}})}\BibitemShut
  {NoStop}%
\bibitem [{\citenamefont {Eliashberg}(1960)}]{Eliashberg1960}%
  \BibitemOpen
  \bibfield  {author} {\bibinfo {author} {\bibfnamefont {G.}~\bibnamefont
  {Eliashberg}},\ }\bibfield  {title} {\enquote {\bibinfo {title} {Interactions
  between electrons and lattice vibrations in a superconductor},}\ }\href
  {https://www.semanticscholar.org/paper/Interactions-between-electrons-and-lattice-in-a-Éliashberg/9985b5de7125e478ff445602cf9b5401473ef911}
  {\bibfield  {journal} {\bibinfo  {journal} {Sov. Phys. JETP}\ }\textbf
  {\bibinfo {volume} {11}},\ \bibinfo {pages} {696} (\bibinfo {year}
  {1960})}\BibitemShut {NoStop}%
\bibitem [{\citenamefont {McMillan}(1968)}]{McMillan1968}%
  \BibitemOpen
  \bibfield  {author} {\bibinfo {author} {\bibfnamefont {W.~L.}\ \bibnamefont
  {McMillan}},\ }\bibfield  {title} {\enquote {\bibinfo {title} {Transition
  temperature of strong-coupled superconductors},}\ }\href {\doibase
  10.1103/PhysRev.167.331} {\bibfield  {journal} {\bibinfo  {journal} {Phys.
  Rev.}\ }\textbf {\bibinfo {volume} {167}},\ \bibinfo {pages} {331--344}
  (\bibinfo {year} {1968})}\BibitemShut {NoStop}%
\bibitem [{\citenamefont {Carbotte}(1990)}]{Carbotte1990}%
  \BibitemOpen
  \bibfield  {author} {\bibinfo {author} {\bibfnamefont {J.~P.}\ \bibnamefont
  {Carbotte}},\ }\bibfield  {title} {\enquote {\bibinfo {title} {Properties of
  boson-exchange superconductors},}\ }\href {\doibase
  10.1103/RevModPhys.62.1027} {\bibfield  {journal} {\bibinfo  {journal} {Rev.
  Mod. Phys.}\ }\textbf {\bibinfo {volume} {62}},\ \bibinfo {pages}
  {1027--1157} (\bibinfo {year} {1990})}\BibitemShut {NoStop}%
\end{thebibliography}%

\end{document}